\documentclass[letter]{aa} 
\usepackage{epsfig}
\usepackage{amsmath}
\usepackage{subfigure}
\usepackage{natbib}
\usepackage{multirow}
\usepackage{color}
\usepackage{ulem}
\usepackage{longtable}
\usepackage{graphicx}
\usepackage{epstopdf}
\usepackage{booktabs}
\usepackage{txfonts}

\newcommand{\cpl}{Chem. Phys. Lett.}

\newcommand{\jms}{J. Mol. Spectr.}
\newcommand{\jmst}{J. Mol. Struct.}

\begin{document}

\title{Discovery of the cyclic C$_5$H radical in TMC-1 \thanks{Based on observations carried out
with the Yebes 40m telescope (projects 19A003, 20A014, 20D023, and 21A011). The 40m radio telescope at Yebes Observatory is operated by the Spanish Geographic Institute (IGN; Ministerio de Transportes, Movilidad y Agenda Urbana).}}

\author{
C.~Cabezas\inst{1},
M.~Ag\'undez\inst{1},
R.~Fuentetaja\inst{1},
Y.~Endo\inst{2},
N.~Marcelino\inst{3,4},
B.~Tercero\inst{3,4},
J.~R.~Pardo\inst{1},
P.~de~Vicente\inst{4}
and
J.~Cernicharo\inst{1}
}

\institute{Grupo de Astrof\'isica Molecular, Instituto de F\'isica Fundamental (IFF-CSIC), C/ Serrano 121, 28006 Madrid, Spain.
\email carlos.cabezas@csic.es; jose.cernicharo@csic.es
\and Department of Applied Chemistry, Science Building II, National Yang Ming Chiao Tung University, 1001 Ta-Hsueh Rd., Hsinchu 300098, Taiwan
\and Observatorio Astron\'omico Nacional (IGN), C/ Alfonso XII, 3, 28014, Madrid, Spain.
\and Centro de Desarrollos Tecnol\'ogicos, Observatorio de Yebes (IGN), 19141 Yebes, Guadalajara, Spain.
}

\date{Received; accepted}

\abstract{Cyclic C$_5$H ($c$-C$_5$H), the radical formed by substituting an ethynyl group CCH for the hydrogen atom in the $c$-C$_3$H radical, has been detected for the first time in the space. The $c$-C$_5$H species is an isomer of the well-known linear radical $l$-C$_5$H and is $\sim$\,6 kcal/mol less stable. A total of 17 rotational transitions were detected for the $c$-C$_5$H species in TMC-1 within the 31.0–50.3 GHz range using the Yebes 40m radio telescope. We derive a column density of (9.0\,$\pm$\,0.9)\,$\times$\,10$^{10}$ cm$^{-2}$ for $c$-C$_5$H. Additionally, we observed 12 lines for $l$-C$_5$H and derive a column density for it of  (1.3\,$\pm$\,0.3)\,$\times$\,10$^{12}$ cm$^{-2}$, which results in an abundance ratio $c$-C$_5$H/$l$-C$_5$H of 0.069. This is in sharp contrast with the value found for the analogue system $c$-C$_3$H/$l$-C$_3$H, whose ratio is 5.5 in TMC-1. We discuss the formation of $c$-C$_5$H and conclude that this radical is probably formed in the reaction of atomic carbon with diacetylene.}


\titlerunning{$c$-C$_5$H in TMC-1}
\authorrunning{Cabezas et al.}

\maketitle

\section{Introduction}

Carbon can be considered the most versatile element for building molecules in the interstellar medium (ISM). In fact, about 80\% of the nearly 260 molecules detected to date in space (CDMS\footnote{https://cdms.astro.uni-koeln.de/}, \citealt{Muller2005}) contain at least one carbon atom, and one-fourth are hydrocarbons. The chemistry of hydrocarbons is dominated to a large extent by highly unsaturated (low H/C ratios) carbon chain molecules such as the C$_n$H family. The hydrocarbon radicals C$_n$H detected in space range from the methylidene radical, CH \citep{Dunham1937}, to C$_8$H \citep{Cernicharo1996,Bell1999}. The smaller ones, such as C$_2$H, C$_3$H, and C$_4$H, are observed in star-forming regions \citep{Tucker1974}, photon-dominated regions \citep{Teyssier2004}, cold dark clouds \citep{Wootten1980,Irvine1981,Thaddeus1985}, translucent molecular clouds \citep{Turner2000}, circumstellar envelopes \citep{Guelin1978,Pardo2007}, and the diffuse medium \citep{Bell1983,Nyman1984}. The larger ones, C$_5$H, C$_6$H, C$_7$H, and C$_8$H, are mainly observed in cold dense molecular clouds \citep{Cernicharo1987,Saito1987,Bell1999,Araki2017} and in the expanding envelope of the carbon-rich star IRC+10216 \citep{Cernicharo1986a,Cernicharo1986b,Suzuki1986,Cernicharo1996,Guelin1996}.

The C$_3$H radical shows a particular behaviour because it exists in two isomeric forms, a cyclic one ($c$-C$_3$H) and a linear one ($l$-C$_3$H), with $c$-C$_3$H lower in energy. Both have been widely observed in the ISM, and in general the $c$-C$_3$H isomer (\citealt{Yamamoto1987,Magnum1990,Turner2000,Cernicharo2000,Zhang2009,Liszt2014}) is found to have a larger abundance than $l$-C$_3$H (\citealt{Thaddeus1985,Turner2000,Pardo2007}). The C$_5$H radical is predicted to adopt up to seven different structures, including linear, cyclic, and bent ones \citep{Crawford1999}. Conversely to C$_3$H, the linear structure, $l$-C$_5$H, is the most stable one. It has been detected in the ISM by \citet{Cernicharo1986a,Cernicharo1986b,Cernicharo1987}. The cyclic isomer, $c$-C$_5$H, which is analogous to the $c$-C$_3$H isomer but with the H atom replaced by a ethynyl (-CCH) group, is the second most stable isomer (6.1 kcal/mol; \citealt{Crawford1999}). As $l$-C$_5$H \citep{McCarthy1999}, $c$-C$_5$H was characterized in the laboratory \citep{Apponi2001}, but it had not yet been detected in the ISM.

In this Letter we report the detection of the $c$-C$_5$H radical towards the cold dark cloud TMC-1. The derived abundance is compared with that of the linear isomer $l$-C$_5$H, and the plausible reactions that could lead to the formation of this species are discussed with the aid of a chemical model.

\section{Observations}

The observational data used in this article consist of spectra of \mbox{TMC-1} taken with the Yebes 40m telescope towards the cyanopolyyne peak of \mbox{TMC-1}, $\alpha_{J2000}=4^{\rm h} 41^{\rm  m} 41.9^{\rm s}$ and $\delta_{J2000}=+25^\circ 41' 27.0''$. The observations are part of the ongoing QUIJOTE\footnote{Q-band Ultrasensitive Inspection Journey to the Obscure TMC-1 Environment.} line survey \citep{Cernicharo2021a} carried out during different observing runs between November 2019 and January 2022. The observations were performed using the frequency-switching mode with a frequency throw of 10 MHz in the very first observing runs, in November 2019 and February 2020, 8 MHz in the observations of January-November 2021, and 10 MHz again in the last observing run that took place between October 2021 and January 2022. The total on-source telescope time is 430 h in each polarization (twice this value after averaging the two polarizations), which can be split into 238 and 192 hours for the 8 MHz and 10 MHz frequency throws. The QUIJOTE line survey uses a 7 mm receiver covering the Q band (31.0-50.3 GHz) with horizontal and vertical polarizations. Receiver temperatures in 2019 and 2020 varied from 22 K at 32 GHz to 42 K at 50 GHz. In 2021, some power adaptation carried out in the down-conversion chains changed the receiver temperatures to 16\,K at 32 GHz and 25\,K at 50 GHz. The backends are 16 Fourier transform spectrometers, which provide a bandwidth of 8\,$\times$\,2.5 GHz in each polarization, thus covering practically the whole Q band, with a spectral resolution of 38.15 kHz. The system is described in detail by \citet{Tercero2021}.

The intensity scale used is antenna temperature, $T_A^*$, which is calibrated using two absorbers at different temperatures and the ATM package \citep{Cernicharo1985,Pardo2001}. Calibration uncertainties were assumed to be 10\% based on the observed repeatability of the line intensities between different observing runs. All data were analysed using the software  GILDAS\footnote{\texttt{http://www.iram.fr/IRAMFR/GILDAS/}}.

\begin{figure}
\centering
\includegraphics[angle=0,width=0.4\textwidth]{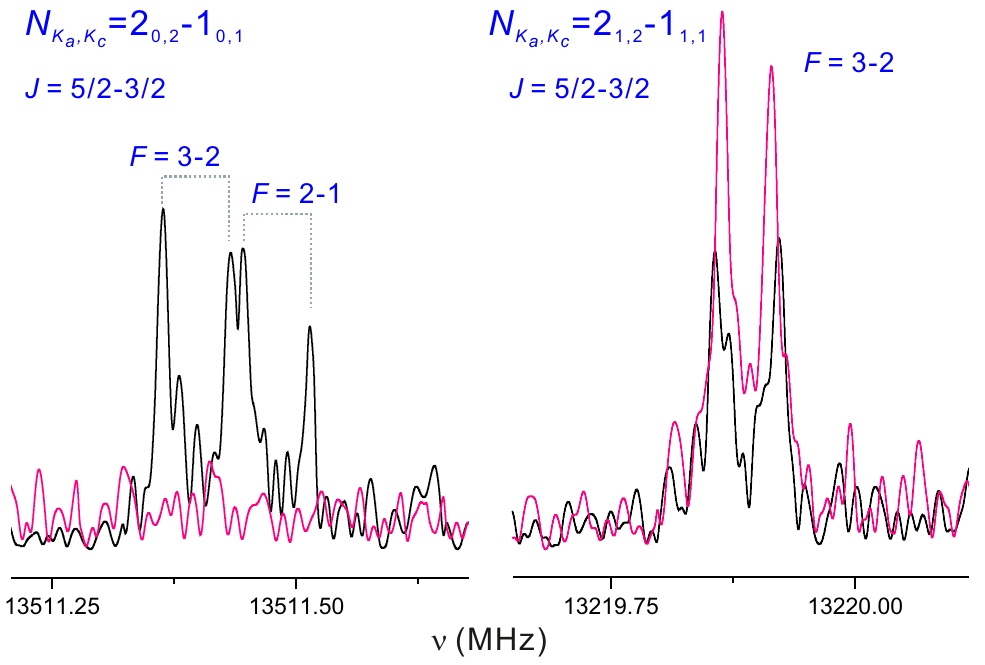}
\caption{FTMW spectra for $c$-C$_5$H. Coloured lines show spectra using argon (magenta) and neon (black) as the carrier gas in the supersonic expansion. Each line is split into two Doppler components because the direction of the supersonic jet expansion is parallel to the standing wave in the Fabry-P\'erot cavity of the spectrometer. The difference in the magnitude of Doppler splitting when different carrier gases -- neon and argon -- are used is related to the velocity of the molecule in the jet; it is reciprocal to the square root of the atomic mass of the carrier gas.} \label{ftmw}
\end{figure}

\section{Results}

\subsection{New laboratory data for $c$-C$_5$H}

\begin{figure*}
\centering
\includegraphics[angle=0,width=\textwidth]{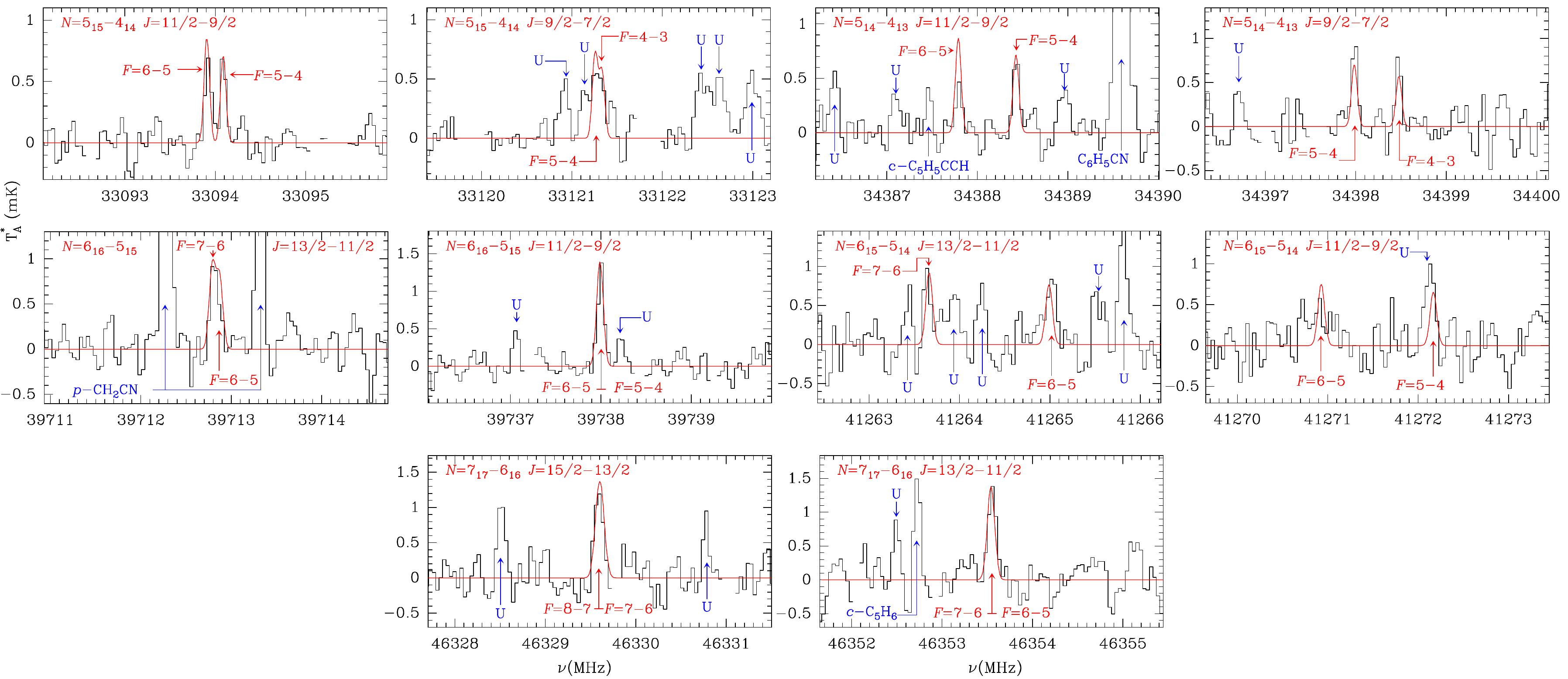}
\caption{Observed lines of $c$-C$_5$H in TMC-1 in the 31.0-50.3 GHz range.
Frequencies and line parameters are given in Table \ref{c-c5h_lines}. Quantum numbers for the observed transitions are indicated in each panel. The red line shows the synthetic spectrum computed for a rotational temperature of 6\,K and a column density of 9.0$\times$10$^{10}$ cm$^{-2}$ (see text). Blanked channels correspond to negative features created in the folding of the frequency-switching data. The label U corresponds to unidentified features above 4$\sigma$.} \label{lines_c-c5h}
\end{figure*}

\citet{Apponi2001} observed the rotational spectrum of the $c$-C$_5$H radical in the laboratory. It has a $^{2}B_2$ electronic ground state with $C_{2v}$ symmetry. Owing to the Bose statistics of the two equivalent off-axis carbon nuclei, only rotational levels with odd $K_a$ occur, and rotational transitions with $K_a$=0 are thus forbidden. \citet{Apponi2001} found two series of lines, with $K_a$=0 and $K_a$=1 values, which were fitted separately. The $K_a$=1 series was ascribed to the ground state of $c$-C$_5$H, while the origin of the $K_a$=0 lines was not totally clear. Our hypothesis is that the $K_a$=0 lines arise from a low frequency vibrationally excited state with different vibrational symmetry (either $B_1$ or $B_2$) from that for the ground state ($A_1$). To gain some insight into this, we carried out Fourier transform microwave (FTMW) spectroscopy experiments using as carrier gases either neon or argon.

The rotational spectrum of the $c$-C$_5$H radical was observed using a Balle-Flygare narrow-band-type FTMW spectrometer operating in the frequency region 4-40 GHz \citep{Endo1994,Cabezas2016}. We observed some of the rotational transitions reported before by \cite{Apponi2001}. The short-lived species $c$-C$_5$H was produced in a supersonic expansion by a pulsed electric discharge of a gas mixture of C$_2$H$_2$ (0.3\%) diluted in neon or argon. The gas mixture was flowed through a pulsed-solenoid valve that is accommodated in the backside of one of the cavity mirrors and aligned parallel to the optical axis of the resonator. A pulse voltage of 1000 V with a duration of 450 $\mu$s was applied between stainless-steel electrodes attached to the exit of the pulsed discharge nozzle, resulting in an electric discharge synchronized with the gas expansion. The resulting products generated in the discharge were then probed by FTMW spectroscopy, which allowed small hyperfine splittings to be resolved.

Previous studies have shown that heavier inert gases have an enhanced cooling efficiency in supersonic expansions due to the larger collision energies they provide \citep{Cabezas2018}. Consequently, an argon-seeded expansion is expected to favour the population of the lower frequency vibrational states, while higher vibrational states will be populated in a neon-seeded expansion. Figure \ref{ftmw} shows the spectra recorded using argon and neon gases. When argon is used, only the $K_a$=1 lines are observed, but in the neon experiments both $K_a$=0 and $K_a$=1 are observed. We note that the intensity of the $K_a$=1 lines decreases when neon is used due to a fraction of its population being distributed between higher vibrationally excited states. In light of our experimental observations, we infer that the $K_a$=0 series of lines come from a low frequency vibrationally excited state, probably an in-plane or out-of-plane bending mode \citep{Crawford1999}.

\subsection{Identification of $c$-C$_5$H in TMC-1}

\begin{table*}
\tiny
    \caption{Observed line parameters for $c$-C$_5$H in TMC-1.}
    \label{c-c5h_lines}
    \centering
    \begin{tabular}{ccccccc}
\hline
\hline
 \multicolumn{3}{c}{Transition} & $\nu_{obs}$~$^a$ & $\int T_A^* dv$~$^b$ & $\Delta v$~$^c$ & $T_A^*$ \\
$N'_{K'_{a,}K'_{c}}$$\leftarrow$$N_{K_{a,}K_{c}}$  & $J'$$\leftarrow$$J$ & $F'$$\leftarrow$$F$  &  (MHz) & (mK\,km\,s$^{-1}$)
& (km\,s$^{-1}$)  & (mK) \\
\hline
$5_{1, 5}$$\leftarrow$$4_{1, 4}$ & 11/2$\leftarrow$9/2  & 6$\leftarrow$5 &  33093.924 & 0.58$\pm$0.09 &   0.79$\pm$ 0.14 &     0.69$\pm$0.11 \\
$5_{1, 5}$$\leftarrow$$4_{1, 4}$ & 11/2$\leftarrow$9/2  & 5$\leftarrow$4 &  33094.088 & 0.50$\pm$0.08 &   0.62$\pm$ 0.14 &     0.75$\pm$0.11 \\
$5_{1, 5}$$\leftarrow$$4_{1, 4}$ &  9/2$\leftarrow$7/2  & 5$\leftarrow$4 &  33121.256 & 0.50$\pm$0.06 &   0.86$\pm$ 0.09 &     0.55$\pm$0.11 \\
$5_{1, 5}$$\leftarrow$$4_{1, 4}$ &  9/2$\leftarrow$7/2  & 4$\leftarrow$3 &  33121.338 & 0.28$\pm$0.05 &   0.64$\pm$ 0.08 &     0.41$\pm$0.11 \\
$5_{1, 4}$$\leftarrow$$4_{1, 3}$ & 11/2$\leftarrow$9/2  & 6$\leftarrow$5 &  34387.798 & 0.33$\pm$0.07 &   0.66$\pm$ 0.17 &     0.46$\pm$0.10 \\
$5_{1, 4}$$\leftarrow$$4_{1, 3}$ & 11/2$\leftarrow$9/2  & 5$\leftarrow$4 &  34388.431 & 0.54$\pm$0.08 &   0.72$\pm$ 0.12 &     0.70$\pm$0.10 \\
$5_{1, 4}$$\leftarrow$$4_{1, 3}$ &  9/2$\leftarrow$7/2  & 5$\leftarrow$4 &  34397.987 & 0.72$\pm$0.12 &   0.68$\pm$ 0.14 &     1.00$\pm$0.20 \\
$5_{1, 4}$$\leftarrow$$4_{1, 3}$ &  9/2$\leftarrow$7/2  & 4$\leftarrow$3 &  34398.472 & 0.55$\pm$0.11 &   0.60$\pm$ 0.14 &     0.86$\pm$0.20 \\
$6_{1, 6}$$\leftarrow$$5_{1, 5}$ & 13/2$\leftarrow$11/2 & 7$\leftarrow$6 &  39712.806 & 0.67$\pm$0.12 &   0.60$\pm$ 0.14 &     1.05$\pm$0.19 \\
$6_{1, 6}$$\leftarrow$$5_{1, 5}$ & 13/2$\leftarrow$11/2 & 6$\leftarrow$5 &  39712.830 & 0.67$\pm$0.18 &   0.76$\pm$ 0.18 &     0.83$\pm$0.19 \\
$6_{1, 6}$$\leftarrow$$5_{1, 5}$ & 11/2$\leftarrow$9/2  & 6$\leftarrow$5 &  \multirow{2}{*}{\bigg\} 39738.002} & \multirow{2}{*}{0.91$\pm$0.09}  &  \multirow{2}{*}{0.64$\pm$ 0.08}  & \multirow{2}{*}{1.34$\pm$0.14}  \\
$6_{1, 6}$$\leftarrow$$5_{1, 5}$ & 11/2$\leftarrow$9/2  & 5$\leftarrow$4 &            &               &                  &                   \\
$6_{1, 5}$$\leftarrow$$5_{1, 4}$ & 13/2$\leftarrow$11/2 & 7$\leftarrow$6 &  41263.639 & 0.65$\pm$0.22 &   0.60$\pm$ 0.22 &     1.02$\pm$0.23 \\
$6_{1, 5}$$\leftarrow$$5_{1, 4}$ & 13/2$\leftarrow$11/2 & 6$\leftarrow$5 &  41265.018 & 0.83$\pm$0.27 &   0.90$\pm$ 0.38 &     0.87$\pm$0.23 \\
$6_{1, 5}$$\leftarrow$$5_{1, 4}$ & 11/2$\leftarrow$9/2  & 6$\leftarrow$5 &  41270.899 & 0.44$\pm$0.12 &   0.77$\pm$ 0.27 &     0.54$\pm$0.23 \\
$6_{1, 5}$$\leftarrow$$5_{1, 4}$ & 11/2$\leftarrow$9/2  & 5$\leftarrow$4 &  41272.163 & 0.45$\pm$0.08~$^d$ &   0.55$\pm$ 0.28 &     0.76$\pm$0.23 \\
$7_{1, 7}$$\leftarrow$$6_{1, 6}$ & 15/2$\leftarrow$13/2 & 8$\leftarrow$7 &  \multirow{2}{*}{\bigg\} 46329.587} & \multirow{2}{*}{0.85$\pm$0.17} & \multirow{2}{*}{0.66$\pm$ 0.13} &  \multirow{2}{*}{1.22$\pm$0.25}    \\
$7_{1, 7}$$\leftarrow$$6_{1, 6}$ & 15/2$\leftarrow$13/2 & 7$\leftarrow$6 &    & & & \\
$7_{1, 7}$$\leftarrow$$6_{1, 6}$ & 13/2$\leftarrow$11/2 & 7$\leftarrow$6 &  \multirow{2}{*}{\bigg\} 46353.556} & \multirow{2}{*}{1.01$\pm$0.22} & \multirow{2}{*}{0.69$\pm$ 0.17} &  \multirow{2}{*}{1.37$\pm$0.25}    \\
$7_{1, 7}$$\leftarrow$$6_{1, 6}$ & 13/2$\leftarrow$11/2 & 6$\leftarrow$5 &    & & & \\
\hline

    \end{tabular}
    \tablefoot{
    \tablefoottext{a}{Observed frequencies towards TMC-1 for which we adopted a v$_{\rm LSR}$ of 5.83 km s$^{-1}$. The frequency uncertainty is 10 kHz.}\tablefoottext{b}{Integrated line intensity in mK\,km\,s$^{-1}$.} \tablefoottext{c}{Full width at half maximum derived by fitting a Gaussian function to the observed line profile (in km\,s$^{-1}$).}\tablefoottext{d}{Line overlaps with an unidentified line.}
}\\
\end{table*}

\begin{table}
\small
\caption{Spectroscopic parameters of $c$-C$_5$H (all in MHz).}
\label{table:constants}
\centering
\begin{tabular}{lcc}
\hline \hline
\multicolumn{1}{c}{Parameter}  & \multicolumn{1}{c}{Lab + TMC-1\,$^a$} & \multicolumn{1}{c}{Laboratory\,$^b$} \\
\hline
$A$                        &  ~      45018.0\,$^c$     &  ~       45018(20)      \\
$B$                        &  ~ 3504.062107(299)\,$^d$ &  ~      3504.0621(3)    \\
$C$                        &  ~    3246.94141(37)      &  ~       3246.9414(4)   \\
$\Delta_N$ $\times$\,10$^3$&  ~      0.2888(52)        &  ~         0.288\,$^e$  \\
$\varepsilon_{aa}$         &  ~    165.1446(216)       &  ~         165.15(2)    \\
$\varepsilon_{bb}$         &  ~       5.1476(56)       &  ~           5.152(6)   \\
$\varepsilon_{cc}$         &  ~     $-$28.1469(38)     &  ~         $-$28.148(4) \\
$a_F$$^{\rm(H)}$           &  ~    $-$11.8018(140)     &  ~         $-$11.79(2)  \\
$T_{aa}$$^{\rm(H)}$        &  ~       6.1686(234)      &  ~           12.13(7)   \\
$T_{bb}$$^{\rm(H)}$        &  ~      $-$0.170(49)        &  ~           5.8(1)     \\
$\sigma$(kHz)              &  ~       9.4              &  ~       -              \\
$N_{lines}$                &  ~       42               &  ~       34             \\
\hline
\end{tabular}
\tablefoot{
\tablefoottext{a}{Merged molecular parameters from a fit to the laboratory and TMC-1 frequencies.}\tablefoottext{b}{Molecular parameters derived by \citet{Apponi2001}. $^c$\,Fixed to the previous determined value.}\tablefoottext{c}{Fixed to the previous determined value.} \tablefoottext{d}{Numbers in parentheses are 1$\sigma$ uncertainties in units of the last digits.}\tablefoottext{e}{Fixed to the C$_5$H$_2$ ring chain value; see \citet{Apponi2001}.}
}\\
\end{table}

\begin{figure}
\centering
\includegraphics[angle=0,width=0.45\textwidth]{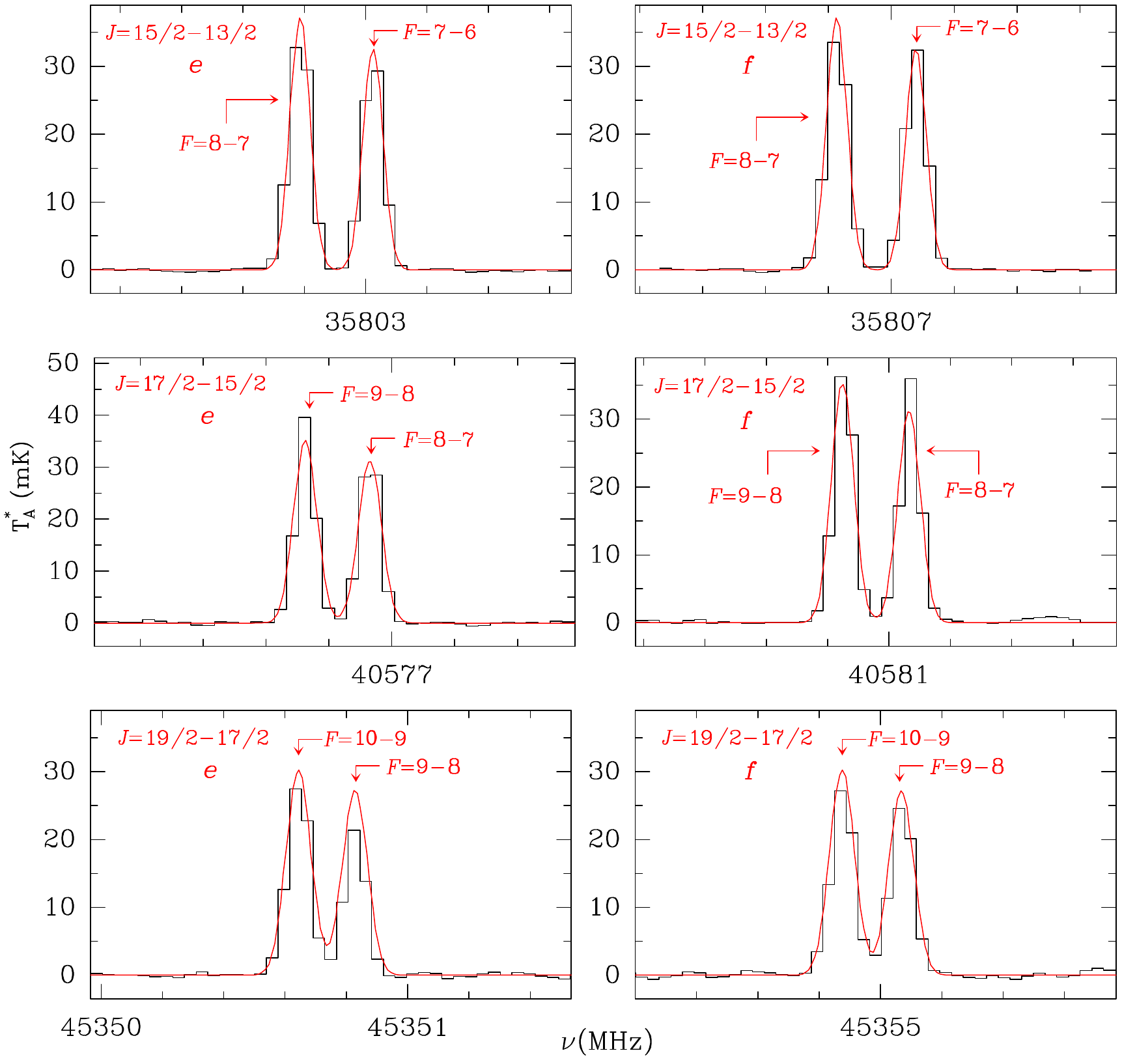}
\caption{Observed lines of C$_5$H in TMC-1 in the 31.0-50.3 GHz range.
Frequencies and line parameters are given in Table \ref{fre_lines}. Quantum numbers for the observed transitions are indicated in each panel. The red line shows the synthetic spectrum computed for a rotational temperature of 6\,K and a column density of 1.3$\times$10$^{12}$ cm$^{-2}$ (see text). Blanked channels correspond to negative features created in the folding of the frequency-switching data.} \label{lines_c5h}
\end{figure}

\begin{table*}
\tiny
    \caption{Observed line parameters for $l$-C$_5$H in TMC-1.}
    \label{fre_lines}
    \centering
    \begin{tabular}{ccccccc}

\hline
\hline
 \multicolumn{3}{c}{Transition} & $\nu_{obs}$~$^a$ & $\int T_A^* dv$~$^b$ & $\Delta v$~$^c$ & $T_A^*$ \\
$J'$$\leftarrow$$J$& $F'$$\leftarrow$$F$ & Parity  &  (MHz)& (mK\,km\,s$^{-1}$) & (km\,s$^{-1}$)  & (mK) \\
\hline
15/2$\leftarrow$13/2 & 8$\leftarrow$7  &$e$  &  35802.786 &    26.79$\pm$0.13 &   0.70$\pm$ 0.01 &    36.19$\pm$0.17 \\
15/2$\leftarrow$13/2 & 7$\leftarrow$6  &$e$  &  35803.024 &    23.10$\pm$0.13 &   0.68$\pm$ 0.01 &    31.81$\pm$0.17 \\
15/2$\leftarrow$13/2 & 8$\leftarrow$7  &$f$  &  35806.828 &    26.36$\pm$0.09 &   0.69$\pm$ 0.01 &    35.73$\pm$0.14 \\
15/2$\leftarrow$13/2 & 7$\leftarrow$6  &$f$  &  35807.078 &    24.00$\pm$0.09 &   0.69$\pm$ 0.01 &    32.69$\pm$0.14 \\
17/2$\leftarrow$15/2 & 9$\leftarrow$8  &$e$  &  40576.724 &    22.85$\pm$0.19 &   0.54$\pm$ 0.01 &    39.42$\pm$0.20 \\
17/2$\leftarrow$15/2 & 8$\leftarrow$7  &$e$  &  40576.931 &    20.37$\pm$0.20 &   0.57$\pm$ 0.01 &    33.53$\pm$0.20 \\
17/2$\leftarrow$15/2 & 9$\leftarrow$8  &$f$  &  40580.854 &    23.55$\pm$0.17 &   0.58$\pm$ 0.01 &    38.45$\pm$0.15 \\
17/2$\leftarrow$15/2 & 8$\leftarrow$7  &$f$  &  40581.071 &    20.82$\pm$0.17 &   0.55$\pm$ 0.01 &    35.62$\pm$0.15 \\
19/2$\leftarrow$17/2 &10$\leftarrow$9  &$e$  &  45350.645 &    18.04$\pm$0.15 &   0.58$\pm$ 0.01 &    29.00$\pm$0.20 \\
19/2$\leftarrow$17/2 & 9$\leftarrow$8  &$e$  &  45350.829 &    12.68$\pm$0.15 &   0.55$\pm$ 0.01 &    21.63$\pm$0.20 \\
19/2$\leftarrow$17/2 &10$\leftarrow$9  &$f$  &  45354.877 &    17.68$\pm$0.20 &   0.60$\pm$ 0.01 &    27.80$\pm$0.35 \\
19/2$\leftarrow$17/2 & 9$\leftarrow$8  &$f$  &  45355.070 &    16.07$\pm$0.20 &   0.59$\pm$ 0.01 &    25.66$\pm$0.35 \\
\hline
    \end{tabular}
    \tablefoot{
\tablefoottext{a}{Observed frequencies towards TMC-1 for which we adopted a v$_{\rm LSR}$ of 5.83 km s$^{-1}$. The frequency uncertainty is 10 kHz.}\tablefoottext{b}{Integrated line intensity in mK\,km\,s$^{-1}$.} \tablefoottext{c}{Full width at half maximum derived by fitting a Gaussian function to the observed line profile (in km\,s$^{-1}$).}
}\\
\end{table*}

Line identification in this work was done using the catalogues MADEX \citep{Cernicharo2012}, CDMS \citep{Muller2005}, and JPL \citep{Pickett1998}. As of May 2022, the MADEX code contained 6434 spectral entries corresponding to the ground and vibrationally excited states, together with the corresponding isotopologues, of 1734 molecules. Once the assignment of all known molecules and their isotopologues is done, QUIJOTE will permit us to search for molecules for which  frequencies are known. Moreover, QUIJOTE also allows us to perform rotational spectroscopy in space of new species for which no previous rotational spectroscopic laboratory information is available, such as HC$_5$NH$^+$ \citep{Marcelino2020}, HC$_3$O$^+$\citep{Cernicharo2020a}, HC$_3$S$^+$ \citep{Cernicharo2021b}, CH$_3$CO$^+$ \citep{Cernicharo2021c}, HCCS$^+$ \citep{Cabezas2022a}, C$_5$H$^+$ \citep{Cernicharo2022}, HC$_7$NH$^+$ \citep{Cabezas2022b}, and HCCNCH$^+$ \citep{Agundez2022}.

Only the $K_a$=1 series of lines of $c$-C$_5$H are expected to be observed in TMC-1. We used the rotational parameters reported by \citet{Apponi2001} to predict, using the SPFIT program \citep{Pickett1991}, the frequency transition lines in the Q band. We used a dipole moment of 3.39 D, following \citet{Crawford1999}. The molecule was implemented in the MADEX code \citep{Cernicharo2012}, which was used to search for the $K_a$=1 lines. A total of 17 lines of $c$-C$_5$H were detected in TMC-1 above the 3$\sigma$ level; they are shown in Fig. \ref{lines_c-c5h}. These 17 lines correspond to 20 hyperfine components, three of which are not resolved, from five $K_a$=1 rotational transitions with $N$=5, 6, and 7. The derived line parameters are given in Table \ref{c-c5h_lines}. A fit to the observed line profiles assuming a source diameter of 40\,$''$ \citep{Fosse2001} provides a rotational temperature of 6.0\,$\pm$\,0.5\,K and a column density of N($c$-C$_5$H)= (9.0\,$\pm$\,0.9)\,$\times$\,10$^{10}$ cm$^{-2}$. The synthetic spectra are compared with observations in Fig. \ref{lines_c-c5h} (red line). No lines with $K_a$=0 were detected, which is in line with our conclusion that those lines belong to a vibrationally excited state that would be under-populated in TMC-1 due to the low kinetic temperature of the cloud.

Using the new observed frequencies in TMC-1 and those measured in the laboratory \citep{Apponi2001}, we have derived, using the SPFIT program \citep{Pickett1991}, a new set of molecular constants for $c$-C$_5$H, which are shown in Table \ref{table:constants}. The new molecular constants are practically identical to those provided by \citet{Apponi2001}, with the exception of the dipole-dipole constants, $T_{aa}$$^{\rm(H)}$ and $T_{bb}$$^{\rm(H)}$. We believe that the definition of these parameters in the fitting code used by \citet{Apponi2001} may be different to that employed by the Pickett program, used in this work, which results in the values being different.

We also observed several lines for the $l$-C$_5$H isomer. A total of 12 hyperfine components for the $J$=15/2-13/2, 17/2-15/2, and 19/2-17/2 transitions, which belong to the $^2$$\Pi_{1/2}$ spin sub-level, were detected (see Fig. \ref{lines_c5h} and Table \ref{fre_lines}) with antenna temperatures around 30 mK. An analysis of the line intensities through a line model fitting procedure \citep{Cernicharo2021d} provides a rotational temperature of 6.0\,$\pm$\,0.5\,K and a column density of N($l$-C$_5$H)=(1.3\,$\pm$\,0.3)\,$\times$\,10$^{12}$ cm$^{-2}$. We assumed a dipole moment of 4.88 D for $l$-C$_5$H \citep{Woon1995}. Therefore, the abundance ratio $c$-C$_5$H/$l$-C$_5$H is 0.069 in TMC-1. This value is far from that found in TMC-1 for the analogue system $c$-C$_3$H/$l$-C$_3$H, whose ratio is 5.5 \citep{Loison2017}.

\section{Chemistry}

In order to shed some light on the formation of \mbox{$c$-C$_5$H} in \mbox{TMC-1}, we carried out gas-phase chemical modelling calculations similar to those presented in \cite{Cabezas2021}. Briefly, we adopted physical conditions typical of cold dark clouds: a volume density of H nuclei of 2\,$\times$10$^4$ cm$^{-3}$, a gas kinetic temperature of 10 K, a visual extinction of 30 mag, a cosmic-ray ionization rate of H$_2$ of 1.3\,$\times$\,10$^{-17}$ s$^{-1}$, and the so-called set of low-metal elemental abundances (e.g., \citealt{Agundez2013}). We employed the chemical network {\small RATE12} from the {\small UMIST} database \citep{McElroy2013}, which has been updated with results from \cite{Lin2013} and expanded with the subset of gas-phase chemical reactions involving C$_3$H and C$_3$H$_2$ isomers revised by \cite{Loison2017}.

\begin{figure}
\centering
\includegraphics[angle=0,width=\columnwidth]{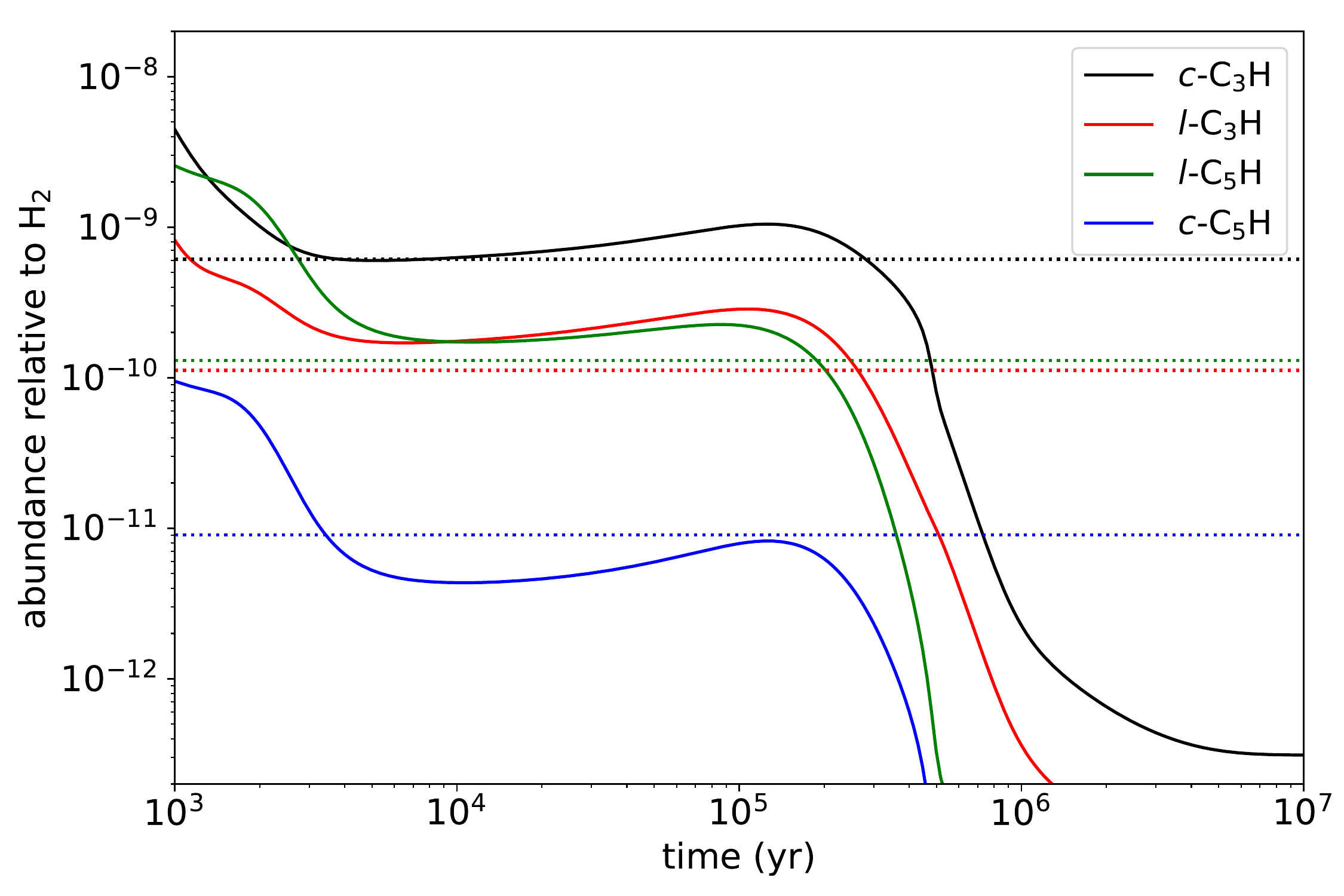}
\caption{Calculated fractional abundances of the cyclic and linear isomers of C$_3$H and C$_5$H as a function of time. The abundances observed in TMC-1 for the isomers of C$_3$H \citep{Loison2017} and C$_5$H (this study) are indicated by dotted horizontal lines.} \label{fig:abun}
\end{figure}

The cyclic isomer of C$_5$H is not included in either the {\small UMIST} \citep{McElroy2013} or {\small KIDA} \citep{Wakelam2015} databases, and information on reactions involving it are lacking in the literature. Our main aim here is thus to explore whether plausible reactions of formation of \mbox{$c$-C$_5$H} can account for the abundance observed in \mbox{TMC-1}. Since \mbox{$c$-C$_5$H} can be viewed as the result of substituting an H atom in $c$-C$_3$H with a C$_2$H group, an obvious reaction of formation of \mbox{$c$-C$_5$H} is
\begin{equation}
\rm c-C_3H + C_2H \rightarrow c-C_5H + H. \label{reac:c3h+c2h}
\end{equation}
We thus included this reaction with a rate coefficient of 3\,$\times$\,10$^{-10}$ cm$^3$ s$^{-1}$, which is similar to rate coefficients measured at low temperature for other reactions of C$_2$H with closed-shell unsaturated hydrocarbons \citep{Chastaing1998}. Another interesting potential source of $c$-C$_5$H is the reaction between atomic carbon and diacetylene,
\begin{subequations}
\begin{align}
\rm C + C_4H_2 & \rightarrow \rm l-C_5H + H, \\
                           & \rightarrow \rm c-C_5H + H,
\end{align}
\end{subequations}
where both the linear and cyclic isomers of C$_5$H are in principle possible, although chemical networks only consider the linear isomer as product \citep{Smith2004,Loison2014}. Reaction (2) is thought to occur rapidly at low temperatures based on experimental studies of reactions of C with closed-shell unsaturated hydrocarbons such as C$_2$H$_2$ \citep{Chastaing2001}. \cite{Takahashi2000} studied this reaction theoretically and concluded that both the linear and cyclic isomers of C$_5$H can be formed without an entrance barrier. On the other hand, calculations by \cite{Sun2008} show that only the linear isomer of C$_5$H should be formed. The different conclusions are likely due to the fact that reaction (2b) was found to be exothermic by 1 kcal mol$^{-1}$ by \cite{Takahashi2000} but endothermic by 0.7 kcal mol$^{-1}$ by \cite{Sun2008}. That is, formation of $c$-C$_5$H is nearly thermo-neutral. Here we assume that reaction (2b) occurs with a branching ratio of just 10\,\%. Finally, we assumed that \mbox{$c$-C$_5$H} reacts fast with neutral atoms (H, C, N, and O) and cations that are known or expected to be abundant in \mbox{TMC-1}, such as C$^+$ and HCO$^+$.

In Fig.~\ref{fig:abun} we show the calculated abundances of the cyclic and linear isomers of C$_3$H and C$_5$H. It is seen that the peak calculated abundances, reached in the 10$^5$-10$^6$ yr time range, agree well with the observed values. If we focus on cyclic C$_5$H, according to the chemical model the main reaction of formation is C + C$_4$H$_2$. That is, if this reaction produces cyclic C$_5$H with only a branching ratio of 10\,\%, then it can by itself explain the abundance of $c$-C$_5$H observed in \mbox{TMC-1}. Further studies to evaluate the branching ratios of the reaction C + C$_4$H$_2$ would allow the chemistry of the linear and cyclic isomers of C$_5$H in cold dark clouds to be better constrained.

\section{Conclusions}

We have reported the first identification of the $c$-C$_5$H radical towards TMC-1. We observed 17 rotational transitions within the 31.0–50.3 GHz range using the Yebes 40m radio telescope. The derived $c$-C$_5$H/$l$-C$_5$H abundance ratio of 0.069 is very different from that found for $c$-C$_3$H/$l$-C$_3$H, whose ratio is 5.5 in TMC-1. A state-of-the-art chemical model reproduces the observed abundance of $c$-C$_5$H and indicates that this radical can probably be formed in the reaction of atomic carbon with diacetylene.

\begin{acknowledgements}

We acknowledge funding support from Spanish Ministerio de Ciencia e Innovaci\'on through grants PID2019-106110GB-I00, PID2019-107115GB-C21, and PID2019-106235GB-I00 and from the European Research Council (ERC Grant 610256: NANOCOSMOS). We
would like to thank Mar\'ia Eugenia Sanz for her useful comments on the laboratory experiments of $c$-C$_5$H radical.\end{acknowledgements}

\end{document}